\documentclass[pra,twocolumn,floatfix]{revtex4}
\usepackage{epsfig}
\usepackage{graphicx}
\usepackage{dcolumn}
\usepackage{amsthm,amsmath}
\usepackage{comment}
\usepackage{float}
\usepackage[colorlinks]{hyperref}
\usepackage{xcolor}

\usepackage{subcaption}
\usepackage[labelformat=simple]{subcaption}

\usepackage{morefloats}

\begin{document}

\title{Magnetic Sublevel Independent Magic and Tune-out Wavelengths of the Alkaline-earth Ions}

\author{Jyoti$^1$}
\author{Harpreet Kaur$^1$}
\author{Bindiya Arora$^1$}
\email{bindiya.phy@gndu.ac.in}
\author{B. K. Sahoo$^2$}

\affiliation{$^1$Department of Physics, Guru Nanak Dev University, Amritsar, Punjab 143005, India}
\affiliation{$^2$Atomic, Molecular and Optical Physics Division, Physical Research Laboratory, Navrangpura, Ahmedabad-380009, India}

\begin{abstract}
Lightshift of a state due to the applied laser in an atomic system vanishes at the tune-out wavelengths ($\lambda_T$s). Similarly, differential 
light shift of a transition vanishes at the magic wavelengths ($\lambda_{magic}$s). In many of the earlier studies, values of the electric dipole
(E1) matrix elements were inferred precisely by combining measurements of $\lambda_{magic}$ with the calculated their values. Similarly, the 
$\lambda_T$ values of an atomic state can be used to infer the E1 matrix element as it involves dynamic electric dipole ($\alpha$) values of only 
one state whereas the $\lambda_{magic}$ values are dealt with $\alpha$ values of two states. However, both the $\lambda_T$ and $\lambda_{magic}$ 
values depend on angular momenta and their magnetic components ($M$) of states. Here, we report the $\lambda_T$ and $\lambda_{magic}$ 
values of many $S_{1/2}$ and $D_{3/2,5/2}$ states, and transitions among these states of the Mg$^{+}$, Ca$^{+}$, Sr$^{+}$ and Ba$^{+}$ ions 
that are independent of $M$- values. Measuring these wavelengths in a special set-up as discussed in the paper, it could be possible to infer
a large number of E1 matrix elements of the above ions accurately. 
\end{abstract}

\maketitle

\section{Introduction}

Singly charged alkaline-earth ions are the most eligible candidates for considering for the high-precision measurements due to several advantages~\cite{zhuang2014active}.
Except Be$^+$ and Mg$^+$, other alkaline-earth ions have two metastable states and most of the transitions among the ground and metastable states 
are accessible by lasers. This is why these ions are considered for carrying out high-precision measurements such as testing Lorentz symmetry 
violations  \cite{kostelecky2018lorentz,Wood19971759,tiecke2014nanophotonic}, parity nonconservation effects \cite{xiaxing1990parity}, non-linear isotope shift effects \cite{PhysRevA.68.022502}, quantum information \cite{PhysRevLett.98.070801,roos2004control} and 
many more including for the optical atomic clock experiments \cite{PhysRevA.72.043404}. One of the major systematics in these measurements is the 
Stark shift due to the employed laser, which depends on the frequency of the laser. The solution to this problem was suggested by Katori et~al.~\cite{katori1999optimal} 
who proposed that the trapping laser can be tuned to wavelengths at which differential ac Stark shifts of the transitions can 
vanish~\cite{katori1999optimal}. These wavelengths were coined as magic wavelengths ($\lambda_{magic}$)s and being popularly used in the optical 
lattice clocks. There are also applications of the magic wavelengths for carrying measurements of atoms trapped inside high-Q cavities in the 
strong-coupling regime~\cite{Mckeever}. In quantum state engineering~\cite{Sackett}, magic wavelengths provide an opportunity to extract accurate 
values of oscillator strengths~\cite{Tang} that are particularly important for the correct stellar modeling and analysis of spectral lines identified 
in the spectra of stars and other heavenly bodies so as to infer fundamental stellar parameters~\cite{ruffoni2014fe,wittkowski2005fundamental}. 

Apart from the magic trapping condition, where light shift of two internal states is identical, another well known limiting case is where light 
shift of one state vanishes. This case is known as tune-out condition~\cite{PhysRevLett.125.023201}. Applications of such tune-out wavelengths  
($\lambda_T$) lie in novel cooling techniques of atoms~\cite{PhysRevLett.103.140401}, selective addressing and manipulation of quantum 
states~\cite{PhysRevLett.115.043003,PhysRevA.73.041405,PhysRevX.9.041014}, precision measurement of atomic structures~\cite{PhysRevLett.109.243004,
PhysRevLett.109.243003,doi:10.1080/00268976.2013.777812,PhysRevLett.115.043004,Kao:17,PhysRevLett.124.203201} and precise estimation of oscillator 
strength ratios~\cite{arora2011tune}. Additionally, tune-out conditions are powerful tools for the evaporative cooling of optical 
lattices~\cite{PhysRevLett.125.023201} and hence, are important for experimental explorations.

In one of the experiments pertaining to magic wavelengths of alkaline-earth ions, Liu et~al. demonstrated the existence of magic wavelengths for a 
single trapped $^{40}$Ca$^{+}$ ion~\cite{Liu} whereas Jiang et~al. evaluated magic wavelengths of Ca$^+$ ions for linearly and circularly polarized 
light using relativistic configuration interaction plus core polarization (RCICP) approach ~\cite{jiang2017magic,jiang2017}. Recently, Chanu et~al. 
proposed a model to trap Ba$^{+}$ ion by inducing an ac Stark shift using $653$  nm linearly polarized laser ~\cite{Chanu_2020}. Kaur et~al. 
reported magic wavelengths for $nS_{1/2}-nP_{1/2,3/2}$ and $nS_{1/2}-mD_{3/2,5/2}$ transitions in alkaline-earth-metal ions using linearly polarized 
light ~\cite{Jasmeet} whereas Jiang et~al. located magic and tune-out wavelengths for Ba$^+$ ion using RCICP approach~\cite{PhysRevA.103.032803}.
Despite having a large number of applications, these magic wavelengths suffer a setback because of their dependency on the magnetic-sublevels ($M$) 
of the atomic systems. Linearly polarized light has been widely used for the trapping of atoms and ions as it is free from the contribution of the 
vector component in the interaction between atomic states and electric fields. However, the magic wavelengths thus identified are again 
magnetic-sublevel dependent for the transitions involving states with angular momenta greater than $1/2$. On the other hand, the implementation 
of circularly polarized light for trapping purposes requires magnetic-sublevel selective trapping. In order to circumvent this $M$-dependency of 
magic wavelengths, a magnetic-sublevel independent strategy for trapping of atoms and ions was proposed by Sukhjit et~al.~\cite{Sukhjit}. Later on, 
Kaur et~al. implemented similar technique to compute magic and tune-out wavelengths independent of magnetic sublevels $M$ for different 
$nS_{1/2}$--$(n-1)D_{3/2,5,2}$ transitions in Ca$^+$, Sr$^+$ and Ba$^+$ ions corresponding to n=$4$ for Ca$^+$, $5$ for Sr$^+$ and $6$ for Ba$^+$ 
ion~\cite{kaur2017annexing}. 

In addition to the applications of $\lambda_{magic}$ in getting rid of differential Stark shift in a transition, they are also being used to infer 
the electric dipole (E1) matrix elements of many allowed transitions in different atomic systems [J. A. Sherman, T. W. Koerber, A. Markhotok, W. Nagourney, and E. N. Fortson
Phys. Rev. Lett. 94, 243001 (2005); B. K. Sahoo, L. W. Wansbeek, K. Jungmann, and R. G. E. Timmermans, Phys. Rev. A 79, 052512 (2009); Liu et al, 
Phys Rev. Lett. 114, 223001 (2015); Jun Jiang, Yun Ma, Xia Wang, Chen-Zhong Dong, and Z. W. Wu, Phys. Rev. A 103, 032803 (2021) etc.]. The
basic procedure of these studies is that the $\lambda_{magic}$ values are calculated by fine-tuning the magnitudes dominantly contributing E1 matrix 
elements to reproduce their measured values. Then, the set of the E1 matrix elements that give rise the best matched $\lambda_{magic}$ values are 
considered as the recommended E1 matrix elements. However, calculations of these $\lambda_{magic}$ values of a transition demand determination 
of dynamic E1 polarizabilities ($\alpha$) of both the states. In view of this, use of $\lambda_T$ values of a given atomic state can be advantageous 
as they involve dynamic $\alpha$ values of only one state. Furthermore, both the $\lambda_T$ and $\lambda_{magic}$ values depend on angular momenta 
and their magnetic components ($M$) of atomic states. This requires evaluation of scalar, vector and tensor components of the $\alpha$ values
for states with angular momenta greater than $1/2$, which is very cumbersome. To circumvent this problem, we present here $M$-sublevel independent
$\lambda_T$ and $\lambda_{magic}$ values of many states and transitions involving a number of $S_{1/2}$ and $D_{3/2,5/2}$ states in the 
alkaline-earth metal ions from Mg$^{+}$ through Ba$^{+}$ that can be inferred to the E1 matrix elements more precisely. We have used the E1 matrix 
elements from an all-order relativistic atomic many-body method to report the $M$-Independent $\lambda_T$ and $\lambda_{magic}$ values to search 
for these values in the experiments, when they are measured precisely the E1 matrix elements need to be fine-tuned in order to minimize their 
uncertainties. It can be achieved by specially setting up the experiment suitably fixing the polarization and quantization angles of the applied 
lasers. To validate our results for the transitions involving high-lying states, we have compared the values of our $\lambda_T$ and 
$\lambda_{magic}$ values for the ground to the metastable states of the considered alkaline-earth ions with the previously reported values. 

The paper is organized as follows: In Sec. \ref{2}, we provide underlying theory and Sec. \ref{Sec3} describes the method of evaluation of the 
calculated quantities. Sec. \ref{3} discusses the obtained results, while concluding the study in Sec. \ref{4}. Unless we have stated explicitly, 
physical quantities are given in atomic units (a.u.).

\section{\label{2} Theory}

\begin{figure}[t!]
\includegraphics[height=7cm,width=9cm]{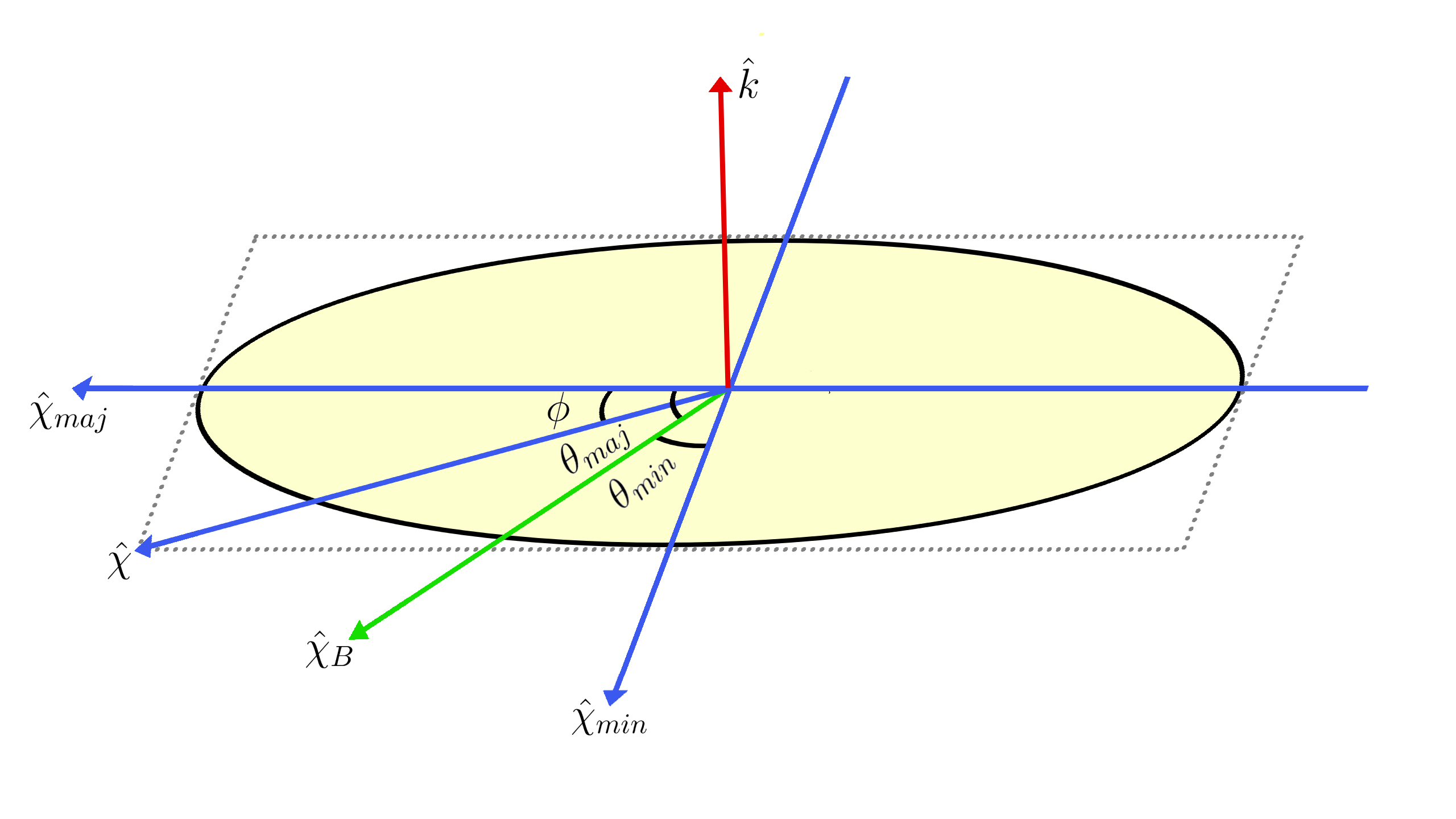}% Here is how to import EPS art
\caption{\label{A}Representation of elliptically polarized laser beam swept out by the laser's polarization vector in one period. $\hat{\chi}$ representing the laser's complex polarization vector and $\hat{k}$ as the laser wave vector perpendicular to  quantization axis $\hat{\chi}_{B}$. The  vectors $\hat{\chi}_{maj}$, $\hat{\chi}_{min}$ and $\hat{k}$  are mutually perpendicular to each other}
\end{figure}

The electric field \textbf{$\mathcal{E}$}(\textbf{$r$},$t)$ associated with a general plane electromagnetic wave can be 
represented in terms of complex polarization vector $\hat{\chi}$ and the real wave vector \textbf{k} by the following expression~\cite{beloy2009theory}
\begin{equation}
\textbf{$\mathcal{E}$}(\textbf{r},t)=\frac{1}{2} \mathcal{E} \hat{\chi} e^{-\iota(\omega t-\textbf{k.r})}+c.c., \label{eq1}
\end{equation}
where $c.c.$ is the complex-conjugate of the preceding term. {Assuming $\hat{\chi}$ to be real and adopting the coordinate system as presented in Fig.~\ref{A}, the polarization vector can be expressed as~\cite{Sukhjit}
\begin{equation}
\hat{\chi}=e^{\iota \sigma} (cos \phi ~\hat{\chi}_{maj}+\iota ~sin \phi ~\hat{\chi}_{min}), \label{eq2}
\end{equation}
where $\hat{\chi}_{maj}$ and $\hat{\chi}_{min}$ denote the real components of the polarization vector $\hat{\chi}$, $\sigma$ is the real quantity denoting the arbitrary phase and $\phi$ is analogous to degree of polarization $A$ such that $A=sin(2\phi)$.}
For linearly polarized light, $\phi=0$ whereas $\phi$ takes the value either $\pi/4$ or $3\pi/4$ for circularly polarized light, which further defines $A=0$ for linearly polarized  and $A=1(-1)$ for right-hand (left-hand) circularly polarized light~\cite{beloy2009theory}.
As shown in the Fig. \ref{A}, this coordinate system follows
 \begin{equation}
cos^2\theta_p=cos^2\phi ~cos^2\theta_{maj}+sin^2\phi ~sin^2\theta_{min} \label{eq3}
\end{equation}
and 
\begin{equation}
\theta_{maj}+\theta_{min}=\frac{\pi}{2}.\label{eq4}
\end{equation}
Here, $\theta_p$ is the angle between quantization axis $\hat{\chi}_{B}$ and direction of polarization vector $\hat{\chi}$ and the parameters 
$\theta_{maj}$ and $\theta_{min}$ are the angles between respective unit vectors and $\hat{\chi_B}$.

When an atomic system is subjected to the above electric field and the magnitude of $\mathcal{E}$ is small, shift in the energy of its $n^{th}$ level
(Stark shift) can be given by
\begin{eqnarray}
 \delta E_n^K \simeq - \frac{1}{2} \alpha_n^K(\omega) |\mathcal{E}|^2, \label{eqpol}
\end{eqnarray}
where $\alpha_n^K(\omega)$ is known as the second-order electric dipole (E1) polarizability and the superscript $K$ denotes angular momentum of the 
state, which can be atomic angular momentum $J$ or hyperfine level angular momentum $F$. Depending upon polarization, dynamic dipole polarizability 
$\alpha_n^K(\omega)$ can be expressed as 
\begin{eqnarray}
\alpha_n^K(\omega)=\alpha_{nS}^{K}(\omega)+\beta(\chi)\frac{M_K}{2K}\alpha^{K}_{nV}(\omega) \nonumber\\
+\gamma(\chi)\frac{3M_K^2-K(K+1)}{K(2K-1)}\alpha_{nT}^K(\omega),\\
\nonumber \label{eqpol}
\end{eqnarray}
where $\alpha_{nS}^{K}$, $\alpha_{nV}^{K}$ and $\alpha_{nT}^K$ are the scalar, vector and tensor components of the polarizability, respectively. In 
the expression  can be defined on the basis of the coordinate system provided in the Fig. \ref{A}. Geometrically, 
values for $\beta(\chi)$ and $\gamma(\chi)$ in their elliptical form are given as~\cite{beloy2009theory,Sukhjit}
\begin{equation}
\beta(\chi)=\iota(\hat{\chi}\times\hat{\chi}^{*}).\hat{\chi}_{B}=Acos\theta_k \label{eq6}
\end{equation}
and
\begin{equation}
	\gamma(\chi)=\frac{1}{2}\left[3(\hat{\chi}^{*}.\hat{\chi}_{B})(\hat{\chi}.\hat{\chi}_{B})-1\right]=\frac{1}{2}\left(3cos^2\theta_p-1\right),
	\label{eq7}
\end{equation}
where $\theta_k$ is the angle between direction of propagation $\textbf{k}$ and ${\hat{\chi}}_{B}$. Substitution of $\beta(\chi)$ and $\gamma(\chi)$ 
from Eq.~\ref{eq6} and ~\ref{eq7} reforms the expression for dipole polarizability to
\begin{eqnarray}
\alpha_n^K(\omega)=\alpha_{nS}^{K}(\omega)+A cos\theta_k \frac{M_K}{2K}\alpha^{K}_{nV}(\omega) \nonumber\\
+\left(\frac{3cos^2\theta_p-1}{2}\right)\frac{3M_K^2-K(K+1)}{K(2K-1)}\alpha_{nT}^K(\omega)
\label{eq8}
\end{eqnarray}
with the azimuthal quantum number $M_K$ of the respective angular momentum $K$. 

Thus, it is obvious from Eq. (\ref{eqpol}) that $\alpha_{n}^K$ values of two states have to be same if we intend to find $\lambda_{magic}$
for the transition involving both the states. Since the above expression for $\alpha_{n}^K$ has $M_K$ dependency, the $\lambda_{magic}$ become 
$M_K$ dependent. In order to remove $M_K$ dependency, one can choose $M_K=0$ sublevels but in the atomic states of the alkaline-earth ions they 
are non-zero while isotopes with integer nuclear spin of the alkaline-earth ions $M_K$s are again non-zero. To address this, a suitable combination 
of the $\beta(\chi)$ and $\gamma(\chi)$ parameters need to be chosen such that $cos\theta_k=0$ and $cos^2\theta_p=\frac{1}{3}$, which are feasible 
to achieve in an experiment by setting $\theta_k, \hat{\chi}_{maj}$ and $\phi$ values as demonstrated in Ref. \cite{Sukhjit}. In such a scenario,
the $\lambda_{magic}$ values can depend on the scalar part only by suppressing the vector and tensor components of $\alpha_{n}^K$; i.e. the net
differential Stark effect of a transition occurring from between the $J$ to $J'$ states will be given by
\begin{equation}
\delta E_{JJ'}=-\frac{1}{2}\left[\alpha_{nS}^{J}(\omega)-\alpha_{nS}^{J'}(\omega)\right]\mathcal{E}^2.
\end{equation}
This has an additional advantage that the differential Stark effects at an arbitrary electric field become independent of choice of atomic or hyperfine 
levels in a given atomic system as the scalar component of $\alpha_{n}^J$ and $\alpha_{n}^F$ are the same. Again, the same choice of $\lambda_{magic}$ 
values will be applicable to both the atomic and hyperfine levels in an high-precision experiment.

\section{\label{Sec3}Method of Evaluation}

Determination of $\alpha_{n}^J$ values require accurate calculations E1 matrix elements. For the computation of E1 matrix elements, we need
accurate atomic wave functions of the alkaline-earth ions. We have employed here a relativistic all-order method to the determine atomic wave 
functions of the considered atomic systems, whose atomic states have closed core configuration with an unpaired electron in the valence orbital. 
Detailed descriptions of our all-order method can be found in Refs.~\cite{blundell1991relativistic,safronova2008all,sahoo2015correlation,
sahoo2015theoretical}, however a brief outline of the same is also provided here for the completeness.

Our all-order method follows the relativistic coupled-cluster (RCC) theory ans\"atz
\begin{equation}
|\psi_v\rangle=e^S|\phi_v\rangle, \label{eqa}
\end{equation}
where $|\phi_v\rangle$ represents the mean-field wave function of the state $v$ and constructed as~\cite{safronova1999relativistic}
\begin{equation}
|\phi_{v}\rangle=a_{v}^{\dagger}|0_{c}\rangle, \label{eqdag}
\end{equation} 
where $|0_{c}\rangle$ represents the Dirac-Hartree-Fock (DHF) wave function of the closed-core. Subscript $v$ represents the valence orbital of the
considered state. In our calculations, we consider only linear terms in the singles and doubles approximation of the RCC theory (SD method)
by expressing~\cite{safronova1999relativistic}
\begin{equation}
|\psi_v\rangle=(1+S_1+S_2+...)|\phi_v\rangle, \label{eq18}
\end{equation}
where $S_1$ and $S_2$ depict terms corresponding to the single and double excitations, respectively, that can further be written in terms of 
second quantization creation and annihilation operators as follows~\cite{iskrenova2007high} 
\begin{eqnarray}
S_1=\sum_{ma} \rho_{ma} a^{\dagger}_m a_a + \sum_{m\neq v}\rho_{mv} a^{\dagger}_m a_v \label{eq19}
\end{eqnarray}
and
\begin{eqnarray}
S_2=\frac{1}{2}\sum_{mnab} \rho_{mnab} a^{\dagger}_m a^{\dagger}_n a_b a_a + \sum_{mna}\rho_{mnva} a^{\dagger}_m a^{\dagger}_n a_a a_v, \ \ \ \  \label{eq20}
\end{eqnarray}
where indices $m$ and $n$ range over all possible virtual orbitals, and indices $a$ and $b$ range over all occupied core orbitals. The 
coefficients $\rho_{ma}$ and $\rho_{mv}$ represent excitation coefficients of the respective single excitations for the core and the valence 
electrons, respectively, whereas $\rho_{mnab}$ and $\rho_{mnva}$ depict double excitation coefficients for the core and the valence electrons 
respectively. These amplitudes are calculated in an iterative procedure~\cite{safronova2007excitation} due to which they include electron 
correlation effects to all-order.

Hence, atomic wave function of the considered states in the alkaline-earth ions are expressed as ~\cite{safronova1999relativistic,kaur2020radiative}:
\begin{eqnarray}\label{eq21}
 |\psi_{v}\rangle_{SD}&&=\left[1+\sum_{ma}\rho_{ma}a^{\dagger}_{m}a_{a}+\frac{1}{2}\sum_{mnab}\rho_{mnab}a^{\dagger}_{m}a^{\dagger}_{n}a_{b}a_{a} \right.
		\nonumber\\
& & \left. +\sum_{m \neq v}\rho_{mv}a^{\dagger}_{m}a_{v}+\sum_{mna}\rho_{mnva}a^{\dagger}_{m}a^{\dagger}_{n}a_{a}a{v}\right]|\phi_{v}\rangle. \ \ \ \
\end{eqnarray} 

To improve the calculations further and understand the importance of contributions from the triple excitations in the RCC theory, we take into account 
important core and valence triple excitations through the perturbative approach over the SD method (SDpT method) by redefining the wave function 
expression as~\cite{safronova1999relativistic}
\begin{eqnarray}\label{eq22}
|\psi_{v}\rangle_{SDpT}& =& |\psi_{v}\rangle_{SD}+\left[\frac{1}{6}\sum_{mnrab}\rho_{mnrvab}a^{\dagger}_{m}a^{\dagger}_{n}a^{\dagger}_{r}a_{b}a_{a}a_{v} \right.
		\nonumber\\
& & \left. +\frac{1}{18}\sum_{mnrabc}\rho_{mnrabc}a^{\dagger}_{m}a^{\dagger}_{n}a^{\dagger}_{r}a_{c}a_{b}a_{a} \right]|\phi_{v}\rangle. \ \ \ \
\end{eqnarray}

After obtaining the wave functions of the interested atomic states, we evaluate the E1 matrix elements between states  
$|\psi_{v}\rangle$ and $|\psi_{w}\rangle$ as~\cite{iskrenova2007high}
\begin{eqnarray}
D_{wv}=\frac{\langle\psi_{w}|D|\psi_{v}\rangle}{\sqrt{\langle\psi_{w}|\psi_{w}\rangle \langle\psi_{v}|\psi_{v}\rangle}}, \label{eq16}
\end{eqnarray}
where $D=-e\Sigma_j \textbf{r}_{j}$ is the E1 operator with $\textbf{r}_j$ being the position of $j^{th}$ electron~\cite{kaur2020radiative}. The 
resulting expression of numerator of Eq.~\ref{eq16} includes the sum of the DHF matrix elements $z_{wv}$, twenty correlation terms of the SD method 
that are linear or quadratic functions of excitation coefficients $\rho_{mv}$, $\rho_{ma}$, $\rho_{mnva}$ and $\rho_{mnab}$, and their core 
counterparts~\cite{safronova2008all}. 

In the sum-over-states approach, expression for the scalar dipole polarizability is given by
\begin{equation}
\alpha_v(\omega) = \frac{2}{3(2J_v+1)}\sum_{v\neq w}\frac{(E_v-E_w)|\langle\psi_v||D||\psi_w\rangle|^2}{(E_v-E_w)^2-\omega^2},
\label{eq23}
\end{equation}
where $\langle\psi_v||D||\psi_w\rangle$ is the reduced matrix element for the transition occurring between the states involving the 
valence orbitals $v$ and $w$. Here, we have dropped the superscript $J$ in the dipole polarizability notation for the brevity. 
For the convenience, we divide the entire contribution to $\alpha_v(\omega)$ in three parts as
\begin{equation}
\alpha_n = \alpha_{n,c} + \alpha_{n,vc} + \alpha_{n,v},\label{eq24}
\end{equation} 
where $c, vc$ and $v$ corresponds to core, valence-core and valence contributions arising due to the correlations among the 
core orbitals, core-valence orbitals and valence-virtual orbitals respectively~\cite{Arora}. Due to very smaller magnitudes, 
the core and core-valence contributions are calculated by using the DHF method. The dominant contributions will arise valence 
from $\alpha_{n,v}$ due to small energy denominators. Again, the high-lying states will not contribute to $\alpha_{n,v}$ owing 
to large energy denominators. Thus, we calculate E1 matrix elements only among the low-lying excited states and refer the 
contributions as `Main'. Contributions from the less contributing high-lying states are referred as `Tail' and are estimated 
again using the DHF method. To reduce the uncertainties in the estimations of Main contributions, we have used experimental 
energies of the states from the National Institute of Science and Technology atomic database (NIST AD)~\cite{ralchenko2005nist}.

\section{\label{3} Results and Discussion}

The precise computation of magic and tune-out wavelengths requires the accurate determination of E1 matrix elements as well as dipole polarizabilities. In our work, we have used E1 matrix elements and energies for different states available on Portal for High-Precision Atomic Data and Computation~\cite{UDportal} and NIST Atomic Spectra Database~\cite{ralchenko2005nist}, respectively.

We have listed resonance transitions, magic wavelengths and their corresponding polarizabilities for magnetic-sublevel independent $nS$--$mD$ transitions for alkaline-earth ions from Mg$^+$ through Ba$^+$ along with their comparison with available literature in the Tables \ref{table1} through \ref{table10}, respectively. The further discussion regarding the magic wavelengths is provided in the subsection \ref{magic} for the considered alkaline-earth ions. Furthermore, we have discussed our results for tune-out wavelengths in the subsection \ref{tune} along with the comparison of our results with respect to the available theoretical data.

\subsection{\label{magic}Magic Wavelengths}
\subsubsection{Mg$^+$}
\begin{figure*}[tb]  
\centering
\begin{subfigure}{0.45\textwidth}
\includegraphics[height=6cm,width=8.5cm]{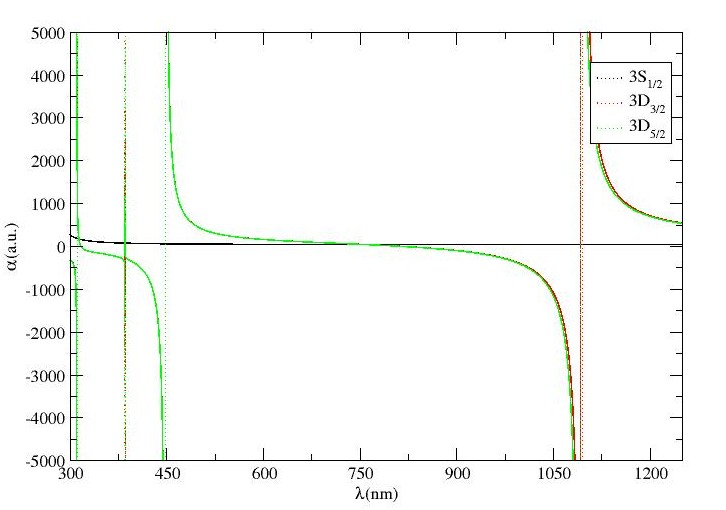}
\caption{$\alpha$ v/s $\lambda$ plot for $3S_{1/2}$ and $3D_{3/2,5/2}$ states of Mg$^{+}$ ion.}
\label{Mg3}
\end{subfigure}
\begin{subfigure}{0.45\textwidth}
\includegraphics[height=6cm,width=8.5cm]{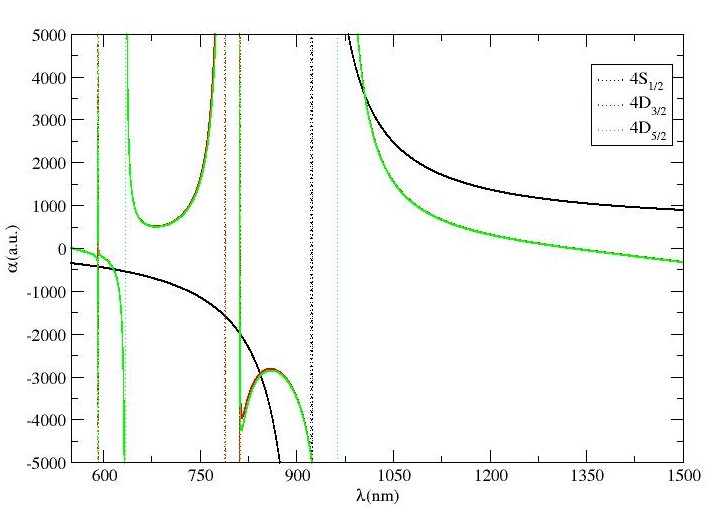}
\caption{$\alpha$ v/s $\lambda$ plot for $4S_{1/2}$--$4D_{3/2,5/2}$ transition in Mg$^{+}$ ion.}
\label{Mg4}
\end{subfigure}
\begin{subfigure}{0.45\textwidth}
\includegraphics[height=6cm,width=8.5cm]{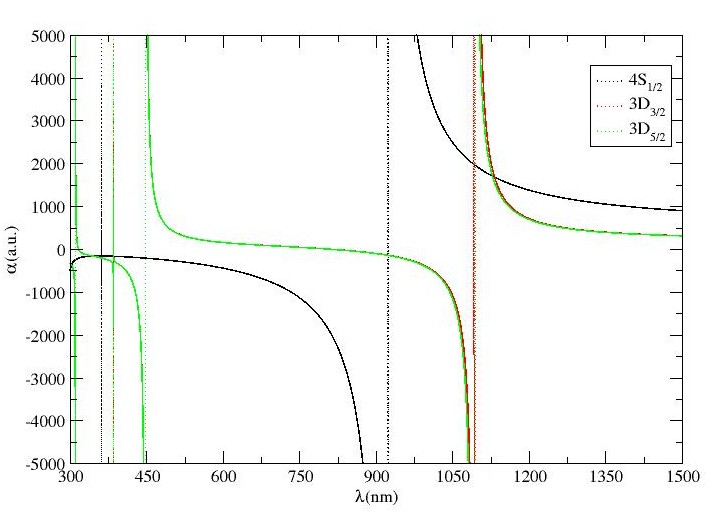}
\caption{$\alpha$ v/s $\lambda$ plot for $4S_{1/2}$--$3D_{3/2,5/2}$ transition in Mg$^{+}$ ion.}
\label{Mg43new}
\end{subfigure}
\begin{subfigure}{0.45\textwidth}
\includegraphics[height=6cm,width=8.5cm]{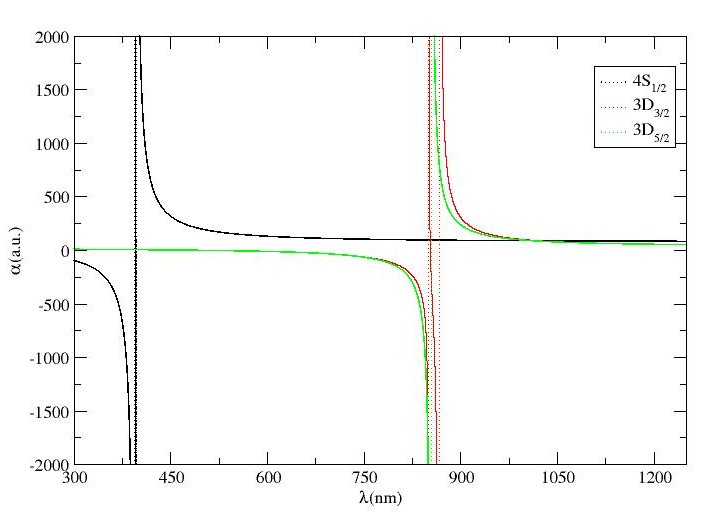}
\caption{$\alpha$ v/s $\lambda$ plot for $4S_{1/2}$--$3D_{3/2,5/2}$ transition in Ca$^{+}$ ion.}
\label{Ca43}
\end{subfigure}
\begin{subfigure}{0.45\textwidth}
\includegraphics[height=6cm,width=8.5cm]{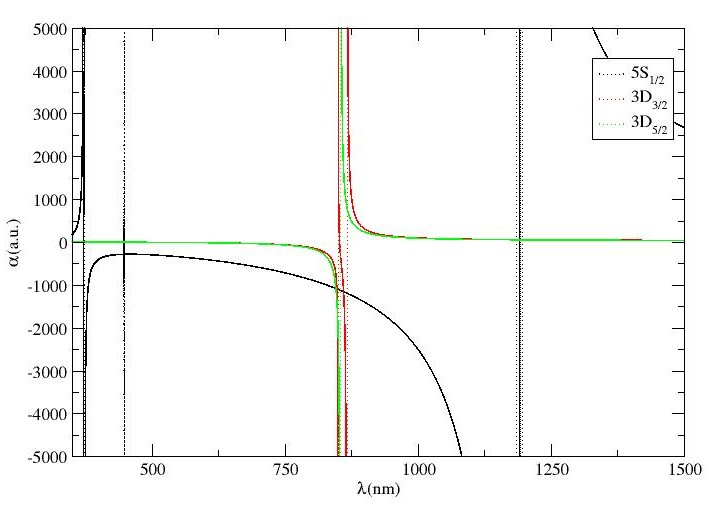}
\caption{$\alpha$ v/s $\lambda$ plot for $5S_{1/2}$--$3D_{3/2,5/2}$ transition in Ca$^{+}$ ion.}
\label{Ca53}
\end{subfigure}
\begin{subfigure}{0.45\textwidth}
\includegraphics[height=6cm,width=8.5cm]{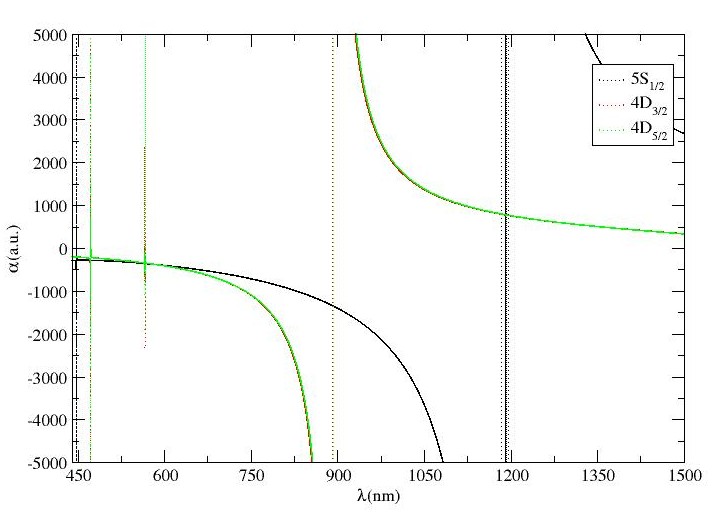}
\caption{$\alpha$ v/s $\lambda$ plot for $5S_{1/2}$--$4D_{3/2,5/2}$ transition in Ca$^{+}$ ion.}
\label{Ca54}
\end{subfigure}
\end{figure*}

\begin{table*}[ht!]
\caption{\label{table1}Magic wavelengths $\lambda_{magic}$ (in nm) with the corresponding polarizability $\alpha_{n}(\omega)$ (in a.u.) for $3S_{1/2}$--$3D_{3/2,5/2}$ transitions in Mg$^{+}$ ion.
	}
	\centering
	\begin{ruledtabular}
\begin{tabular}{cccccccccc}
\multicolumn{4}{c}{$3S_{1/2}-3D_{3/2}$}  & & \multicolumn{4}{c}{$3S_{1/2}-3D_{5/2}$}\\
Resonance & $\lambda_{res}$ & $\lambda_{magic}$ & $\alpha_{magic}$ & & Resonance & $\lambda_{res}$ & $\lambda_{magic}$ & $\alpha_{magic}$\\
\hline
$3D_{3/2}\rightarrow 6P_{3/2}$ & $292.92$ & & & & $3D_{5/2}\rightarrow 5F_{5/2}$ & $310.56$ & & &\\
&& $313.89$ & $168.72$ & & & & $313.86$ & $168.85$\\
$3D_{3/2}\rightarrow 5P_{1/2}$ & $385.15$ & & & & $3D_{5/2}\rightarrow 5P_{3/2}$ & $384.920$
& & \\
&& $385.30$ & $73.51$ & & & & $385.10$ & $73.59$\\
&& $757.79$ & $40.43$ & & & & & &\\
$3D_{3/2}\rightarrow 5P_{3/2}$ & $1091.83$ & & & & $3D_{5/2}\rightarrow 4F_{5/2}$ & $448.24$ & &\\
&& $1092.44$ & $ 37.41$ & & & & $756.72$ & $40.45$\\
$3D_{3/2}\rightarrow 4P_{1/2}$ & $1095.48$ & & & & $3D_{5/2}\rightarrow 4P_{3/2}$ & $1091.72$ & &\\
&&&&&&&&\\
\hline
\multicolumn{4}{c}{$4S_{1/2}-3D_{3/2}$}  & & \multicolumn{4}{c}{$4S_{1/2}-3D_{5/2}$}\\
\hline
&&&&&&&&\\
$4S_{1/2}\rightarrow 5P_{3/2}$ & $361.48$ & & & & $4S_{1/2}\rightarrow 5F_{7/2}$ & $310.56$ & & &\\
&& $361.26$ & $-202.05$ & & & & $344.87$ & $-160.00$\\
&&&&&&&$361.63$ & $-202.81$\\
$4S_{1/2}\rightarrow 5P_{1/2}$ & $361.66$ & & & & $4S_{1/2}\rightarrow 5P_{3/2}$ & $361.48$
& & \\
&& $361.63$ & $-203.81$ & & & & $361.62$ & $-203.15$\\
$4S_{1/2}\rightarrow 4P_{3/2}$ & $922.08$ & & & & $4S_{1/2}\rightarrow 5P_{1/2}$ & $361.66$ & &\\
&& $923.81$ & $-140.74$ & & & & &\\
$4S_{1/2}\rightarrow 4P_{1/2}$ & $924.68$ & & & & $3D_{5/2}\rightarrow 5P_{3/2}$ & $384.93$ & &\\ 
&&&&&&& $385.41$ & $-163.67$\\
$3D_{3/2}\rightarrow 4P_{3/2}$ & $1091.83$ & & & & $3D_{5/2}\rightarrow 4F_{5/2}$ & $448.24$ & &\\
&& $1092.38$ & $1976.85$ &&&&&&\\
$3D_{3/2}\rightarrow 4P_{1/2}$ & $1095.48$ & & & & $4S_{1/2}\rightarrow 4P_{3/2}$ & $922.08$ & &\\
& & $1132.53$ & $1681.41$ & & && $923.81$ & $-144.72$\\
&&&&& $4S_{1/2}\rightarrow 4P_{1/2}$ & $924.68$ & &\\
&&&&& $ 3D_{5/2}\rightarrow 4P_{3/2}$ & $1091.72$ & &\\
&&&&&&& $1128.42$ & $1706.33$\\
&&&&&&&&\\
\hline
\multicolumn{4}{c}{$4S_{1/2}-4D_{3/2}$}  & & \multicolumn{4}{c}{$4S_{1/2}-4D_{5/2}$}\\
\hline
&&&&&&&&\\
$4D_{3/2}\rightarrow 7F_{5/2}$ & $526.58$ & & & & $4D_{5/2}\rightarrow 7F_{5/2,7/2}$ & $526.57$ & & &\\
& & $591.48$ & $-422.01$ & & & & $591.34$ & $-421.69$\\
$4D_{3/2}\rightarrow 7P_{3/2}$ & $591.83$ & & & & $4D_{5/2}\rightarrow 7P_{3/2}$ & $591.81$ & & &\\
&& $591.86$ & $-422.86$ & & & & &\\
$4D_{3/2}\rightarrow 7P_{1/2}$ & $591.98$ & & & & & & & \\
&& $616.11$ & $-482.44$ & & & & $616.02$ & $-482.19$\\
$4D_{3/2}\rightarrow 6F_{5/2}$ & $634.87$ & & & & $4D_{5/2}\rightarrow 6F_{7/2}$ & $634.85$ & & &\\
$4P_{1/2}\rightarrow 4D_{3/2}$ & $787.92$ & & & & & & & &\\
&& $789.50$ & $-1578.93$ & & & & &\\
$4P_{3/2}\rightarrow 4D_{3/2}$ & $789.82$ & & & & $4P_{3/2}\rightarrow 4D_{5/2}$ & $789.85$ & &\\
$4D_{3/2}\rightarrow 6P_{3/2}$ & $811.78$ & & & & $4D_{5/2}\rightarrow 6P_{3/2}$ & $811.75$ & &\\
&& $811.82$ & $-1973.57$ & & & & $812.12$ & $-1979.96$\\
&& & & & & & $844.63$ & $-2964.83$\\
$4D_{3/2}\rightarrow 6P_{1/2}$ & $812.83$ & & & & & & &\\
&& $812.65$ & $-1991.34$ & & & & &\\
&& $843.61$ & $-2921.35$ & & & & &\\
$4S_{1/2}\rightarrow 4P_{3/2}$ & $922.08$ & & & & $4S_{1/2}\rightarrow 4P_{3/2}$ & $922.08$ & &\\
&& $923.84$ & $-4957.89$ & & & & $923.84$ & $-4975.83$\\
$4S_{1/2}\rightarrow 4P_{1/2}$ & $924.68$ & & & & $4S_{1/2}\rightarrow 4P_{1/2}$ & $924.68$ & &\\
$4D_{3/2}\rightarrow 5F_{5/2}$ & $963.51$ & & & & $4D_{5/2}\rightarrow 5F_{7/2}$ & $963.45$ & &\\
&& $1006.101$ & $3585.53$ & & & & $1005.86$ & $3594.76$\\
		\end{tabular}
		\end{ruledtabular}
		\end{table*} 
In Table~\ref{table1}, we have tabulated our results for magic wavelengths and their corresponding dipole polarizabilities for $(3,4)S_{1/2}$--$3D_{3/2,5/2}$ and $4S_{1/2}$--$4D_{3/2}$ transitions. Fig. \ref{Mg3} demonstrates scalar dipole polarizabilities of $3S_{1/2}$ and $3D_{3/2,5/2}$ states of Mg$^+$ ion with respect to wavelength of the external field. It can be perceived from the figure that a number of magic wavelengths at the crossings of the scalar polarizabilities' curves of the corresponding state have been predicted for the transition. As can be seen from the Table \ref{table1}, a total of $4$ magic wavelengths have been found for $3S$--$3D_{3/2}$ transition, whereas $3S$--$3D_{5/2}$ transition shows a total of $3$ magic wavelengths in the range $300-1250$ nm, out of which no magic wavelength is found to exist in visible spectrum. However, all the magic wavelengths enlisted in Table \ref{table1} support red-detuned trap.

Fig. \ref{Mg4} represents the plot of scalar dipole polarizabilities of $4S$ and $4D_{3/2,5/2}$ states against wavelength of the external field. It can also be assessed from Table~\ref{table1} that there exists a total of nine magic wavelengths in the considered wavelength range for $4S$--$4D_{3/2}$ transition, whereas only five magic wavelengths are spotted for $4S$--$4D_{5/2}$ transition. However, in both the cases, all the magic wavelengths except those around $616$ nm, $844$ nm and $1006$ nm are close to resonance, thereby making them unsuitable for further use. However, out of these three values, $\lambda_{magic}$ at $616$ nm lies in the visible region and is far-detuned with considerable deep potential. Hence, we recommend this magic wavelength for trapping of Mg$^+$ ion for both $4S$--$4D_{3/2,5/2}$ transitions for further experimentations in optical clock applications.

Fig. \ref{Mg43new} demonstrates the magic wavelengths for M$_J$ independent scheme for $4S$--$3D_{3/2,5/2}$ transitions for Mg$^+$ ion along with their corresponding scalar dynamic polarizabilities. According to Table \ref{table1}, it can be realized that none of the magic wavelengths for these transitions lies within the visible spectrum of electromagnetic radiations. However, all of these magic wavelengths support red-detuned trap, except $1132.53$ nm and $1128.42$ nm for $4S$-$3D_{3/2}$ and $4S$--$3D_{5/2}$ transitions, respectively, support far blue-detuned traps and are found to be useful for experimental demonstrations.

\subsubsection{Ca$^+$}
\begin{table*}[ht!]
\caption{\label{table4}Magic wavelengths $\lambda_{magic}$ (in nm) with the corresponding polarizability $\alpha_{n}(\omega)$ (in a.u.) for $4S_{1/2}$--$3D_{3/2,5/2}$ transitions in Ca$^{+}$ ion and their comparison with available literature.
	}
	\begin{ruledtabular}
\begin{tabular}{cccccccccc}
\multicolumn{4}{c}{$4S_{1/2}-3D_{3/2}$}  & & \multicolumn{4}{c}{$4S_{1/2}-3D_{5/2}$}\\
Resonance & $\lambda_{res}$ & $\lambda_{magic}$ & $\alpha_{magic}$ & & Resonance & $\lambda_{res}$ & $\lambda_{magic}$ & $\alpha_{magic}$\\
\hline
$4S_{1/2}\rightarrow 4P_{3/2}$ & $393.48$ & & & & $4S_{1/2}\rightarrow 4P_{3/2}$ & $393.47$ & & &\\
&& $395.80$ & $5.57$ & & & & $395.80$ & $5.57$\\
&&$395.82(3)$~\cite{kaur2017annexing} & $4.90$~\cite{kaur2017annexing} &&&& $395.82(2)$~\cite{kaur2017annexing} & $4.20$~\cite{kaur2017annexing}\\
$4S_{1/2}\rightarrow 4P_{1/2}$ & $396.96$ & & & & $4S_{1/2}\rightarrow 4P_{1/2}$ & $396.96$
& & \\
$3D_{3/2}\rightarrow 4P_{3/2}$ & $850.04$ & & & & $3D_{5/2}\rightarrow 4P_{3/2}$ & $854.44$ & &\\
&& $852.42$ & $95.67$ & & & & $1011.90$ & $88.89$\\
&& $852.45(2)$~\cite{kaur2017annexing} & $4.20$~\cite{kaur2017annexing} &&&& $1014.10(3)$~\cite{kaur2017annexing} & $89.01$~\cite{kaur2017annexing} \\
$3D_{3/2}\rightarrow 4P_{1/2}$ & $866.45$ & & & & & & &\\
&& $1028.97$ & $88.39$ &&&&&&\\
&& $1029.7(2)$~\cite{kaur2017annexing} & $88.55$~\cite{kaur2017annexing} &&&&&&\\
&&&&&&&&\\
\hline
\multicolumn{4}{c}{$5S_{1/2}-3D_{3/2}$}  & & \multicolumn{4}{c}{$5S_{1/2}-3D_{5/2}$}\\
\hline
&&&&&&&&\\
$4P_{1/2}\rightarrow 5S_{1/2}$ & $370.71$ & & & & $4P_{1/2}\rightarrow 5S_{1/2}$ & $370.71$  & & &\\
&& $371.76$ & $6.66$ & & & & $371.76$ & $6.67$\\
$4P_{3/2}\rightarrow 5S_{1/2}$ & $373.80$ & & & & $4P_{3/2}\rightarrow 5S_{1/2}$ & $373.80$
& & \\
$5S_{1/2}\rightarrow 6P_{3/2}$ & $447.33$ & & & & $5S_{1/2}\rightarrow 6P_{3/2}$ & $447.33$ & &\\
&& $447.39$ & $2.95$ & & & & $447.39$ & $2.92$\\
$5S_{1/2}\rightarrow 6P_{1/2}$ & $448.07$ & & && $5S_{1/2}\rightarrow 6P_{1/2}$ & $448.07$ &&\\
&& $448.09$ & $2.91$ & & & & $448.09$ & $2.88$\\
&& $847.69$ & $-1089.37$ & & & & $845.78$ & $-1079.82$\\
$3D_{3/2}\rightarrow 4P_{3/2}$ & $850.04$ & & & & $3D_{5/2}\rightarrow 4P_{3/2}$ & $854.44$ & &\\
&& $860.22$ & $-1154.74$ & & & & &\\
$3D_{3/2}\rightarrow 4P_{1/2}$ & $866.45$ & & & & &\\
$5S_{1/2}\rightarrow 5P_{3/2}$ & $1184.22$ & & & & $5S_{1/2}\rightarrow 5P_{3/2}$ & $1184.22$\\
&& $1191.59$ & $58.93$ & & & & $1191.59$ & $56.82$\\
$5S_{1/2}\rightarrow 5P_{1/2}$ & $1195.30$ & & & &$5S_{1/2}\rightarrow 5P_{1/2}$ & $1195.30$&&&\\
&&&&&&&&\\
\hline
\multicolumn{4}{c}{$5S_{1/2}-4D_{3/2}$}  & & \multicolumn{4}{c}{$5S_{1/2}-4D_{5/2}$}\\
\hline
&&&&&&&&\\
$5S_{1/2}\rightarrow 6P_{3/2}$ & $447.33$ & & & & $5S_{1/2}\rightarrow 6P_{3/2}$ & $447.34$ & & &\\
&& $447.52$ & $-204.20$ & & & & $447.52$ & $-205.30$\\
$5S_{1/2}\rightarrow 6P_{1/2}$ & $448.07$ & & & & $5S_{1/2}\rightarrow 6P_{1/2}$ & $448.07$
& & \\
& & $448.16$ & $-204.70$ & & & & $448.16$ & $-205.79$\\
& & $471.22$ & $-279.71$ & & & & $471.67$ & $-279.85$\\
$4D_{3/2}\rightarrow 5F_{5/2}$ & $471.81$ & & & & $4D_{5/2}\rightarrow 5F_{5/2,7/2}$ & $472.23$ & &\\
&& $565.51$ & $-355.55$ & & & & $566.03$ & $-356.17$\\
$4D_{3/2}\rightarrow 6P_{3/2}$ & $565.53$ & & && $4D_{5/2}\rightarrow 6P_{3/2}$ & $566.15$ &&\\
&& $566.67$ & $-356.93$ &&&&&&\\
$4D_{3/2}\rightarrow 6P_{1/2}$ & $566.71$ & & & & & &\\
$4D_{3/2}\rightarrow 4F_{5/2}$ & $891.45$ & & & &$4D_{5/2}\rightarrow 4F_{7/2}$ & $892.98$\\
$5S_{1/2}\rightarrow 5P_{3/2}$ & $1184.22$ & & & & $5S_{1/2}\rightarrow 5P_{3/2}$ & $1184.22$\\
&& $1191.56$ & $776.08$ & & & & $1191.56$ & $779.87$\\
$5S_{1/2}\rightarrow 5P_{1/2}$ & $1195.30$ & & & &$5S_{1/2}\rightarrow 5P_{1/2}$ & $1195.302$&&&\\
		\end{tabular}
		\end{ruledtabular}
		\end{table*}
		 
We have considered $4S$--$3D_{3/2,5/2}$ and $5S$--$(4,3)D_{3/2,5/2}$ transitions for locating the magic wavelengths in Ca$^+$ ion. We have tabulated magic wavelengths for these transitions along with the comparison of $\lambda_{magic}$s with the only available results for $4S$--$3D_{3/2,5/2}$ in Table \ref{table4}. Also, we have plotted scalar dipole polarizabilities against wavelengths for these transitions in Figs. \ref{Ca43}, \ref{Ca53} and \ref{Ca54} correspondingly. According to Table \ref{table4}, it is ascertain that subsequently three and two magic wavelengths exist between $393$ nm and $1030$ nm for $4S$--$3D_{3/2,5/2}$ transitions. In both cases, except $1029.97$ nm and $1011.90$ nm magic wavelengths, that are far-detuned, all other magic wavelengths are close to resonances and are not suitable for laser trapping.

During analysis, six and five magic wavelengths are located for $5S$--$(3,4)D_{3/2}$ and $5S$--$(3,4)D_{5/2}$ transitions, respectively. It is also analyzed that all the magic wavelengths are approximately same for both $5S$--$4D_{3/2}$ and $5S$--$4D_{5/2}$ transitions. Moreover, $\lambda_{magic}$s around $845$ nm, $847$ nm and $860$ nm share deep trapping potential for blue-detuned traps and hence, are further recommended for configuring feasible traps. $\lambda_{magic}$ at $1191.56$ nm, identified in infrared region for both $5S$-$4D_{3/2,5/2}$ transitions, is the only magic wavelength that supports red-detuned trap. Besides, the polarizability for this wavelength is sufficient enough for creating an ion trap at reasonable laser power. To validate our results, we have also compared our results with the results provided only for $4S$--$3D_{3/2,5/2}$ in Ref.~\cite{kaur2017annexing}, and noticed that the results for these transitions are in good agreement with only less than $1\%$ variation w.r.t.  obtained results.

\subsubsection{Sr$^+$}	
\begin{table*}[ht!]
\caption{\label{table7}Magic wavelengths $\lambda_{magic}$ (in nm) with the corresponding polarizability $\alpha_{n}(\omega)$ (in a.u.) along with their comparison with available literature for $5S_{1/2}$--$4D_{3/2,5/2}$ transitions in Sr$^{+}$ ion.
	}
	\begin{ruledtabular}
\begin{tabular}{cccccccccc}
\multicolumn{4}{c}{$5S_{1/2}-4D_{3/2}$}  & & \multicolumn{4}{c}{$5S_{1/2}-4D_{5/2}$}\\
Resonance & $\lambda_{res}$ & $\lambda_{magic}$ & $\alpha_{magic}$ & & Resonance & $\lambda_{res}$ & $\lambda_{magic}$ & $\alpha_{magic}$\\
\hline
$5S_{1/2}\rightarrow 5P_{3/2}$ & $407.89$ & & & & $5S_{1/2}\rightarrow 5P_{3/2}$ & $407.89$ & & &\\
&& $417.00$ & $15.28$ & & & & $417.00$ & $15.18$\\
&& $416.9(3)$~\cite{kaur2017annexing} & $14.47$~\cite{kaur2017annexing} &&&& $416.9(3)$~\cite{kaur2017annexing} & $13.3$~\cite{kaur2017annexing} \\
$5S_{1/2}\rightarrow 5P_{1/2}$ & $421.67$ & & & & $5S_{1/2}\rightarrow 5P_{1/2}$ & $421.67$
& & \\
$4D_{3/2}\rightarrow 5P_{3/2}$ & $1003.94$ & & & & $4D_{5/2}\rightarrow 5P_{3/2}$ & $1003.01$ & &\\
&& $1014.68$ & $108.70$ & & & & & &\\
&& $1014.6(2)$~\cite{kaur2017annexing} & $108.35$~\cite{kaur2017annexing} &&&&&&\\
$4D_{3/2}\rightarrow 5P_{1/2}$ & $1091.79$ & & && && &&\\
&&&&&&&&\\
\hline
\multicolumn{4}{c}{$6S_{1/2}-4D_{3/2}$}  & & \multicolumn{4}{c}{$6S_{1/2}-4D_{5/2}$}\\
\hline
&&&&&&&&\\
$5P_{1/2}\rightarrow 6S_{1/2}$ & $416.27$ & & & & $5P_{1/2}\rightarrow 6S_{1/2}$ & $416.30$ & & &\\
&& $421.47$ & $14.98$ & & & & $421.47$ & $14.85$\\
$5P_{3/2}\rightarrow 6S_{1/2}$ & $430.67$ & & & & $5P_{3/2}\rightarrow 6S_{1/2}$ & $430.67$
& & \\
$6S_{1/2}\rightarrow 7P_{3/2}$ & $474.37$ & & & & $6S_{1/2}\rightarrow 7P_{3/2}$ & $474.37$ & &\\
&& $474.61$ & $11.35$ & & & & $474.61$ & $10.95$\\
$6S_{1/2}\rightarrow 7P_{1/2}$ & $477.49$ & & && $6S_{1/2}\rightarrow 7P_{1/2}$ & $477.49$ &&\\
&& $477.55$ & $11.14$ & & & & $477.56$ & $10.72$\\
&& $1002.40$ & $-2470.01$ & & & & $1025.19$ & $-2857.98$\\
$4D_{3/2}\rightarrow 5P_{3/2}$ & $1003.94$ & & && $4D_{5/2}\rightarrow 5P_{3/2}$ & $1033.01$ & &&\\
&& $1087.35$ & $-4653.36$ & & & & & &\\
$4D_{3/2}\rightarrow 5P_{1/2}$ & $1091.79$ & & && $6S_{1/2}\rightarrow 6P_{3/2}$ & $1201.73$ &&\\
$6S_{1/2}\rightarrow 6P_{3/2}$ & $1201.73$ & & && && $1230.05$ & $170.22$\\
&& $1230.02$ & $223.42$ & & & & & &\\
$6S_{1/2}\rightarrow 6P_{1/2}$ & $1244.84$ & & && $6S_{1/2}\rightarrow 6P_{1/2}$ & $1244.84$  &&\\
&&&&&&&&\\
\hline
\multicolumn{4}{c}{$6S_{1/2}-5D_{3/2}$}  & & \multicolumn{4}{c}{$6S_{1/2}-5D_{5/2}$}\\
\hline
&&&&&&&&\\
$5D_{3/2}\rightarrow 5F_{5/2}$ & $562.45$ & & & & $5D_{5/2}\rightarrow 5F_{5/2}$ & $565.20$ & & &\\
&& $643.87$ & $-528.64$ & & & & $647.41$ & $-534.62$\\
$5D_{3/2}\rightarrow 7P_{3/2}$ & $643.88$ & & & & $5D_{5/2}\rightarrow 7P_{3/2}$ & $647.49$
& & \\
&& $649.49$ & $-538.20$ & & & & &\\
$5D_{3/2}\rightarrow 7P_{1/2}$ & $649.65$ & & && & &&\\
$6S_{1/2}\rightarrow 6P_{3/2}$ & $1201.73$ & & && $6S_{1/2}\rightarrow 6P_{3/2}$ & $1201.73$ & &&\\
&& $1233.61$ & $-6755.64$ & & & & $1233.06$ & $-5620.27$\\
$6S_{1/2}\rightarrow 6P_{1/2}$ & $1244.84$ & & && $6S_{1/2}\rightarrow 6P_{1/2}$ & $1244.84$ &&\\
$5D_{3/2}\rightarrow 4F_{5/2}$ & $1297.85$ & & && $5D_{5/2}\rightarrow 4F_{5/2}$ & $1312.62$ & &\\
&& $1411.88$ & $4381.56$ & & & & & &\\
&&&&&$5D_{5/2}\rightarrow 4F_{7/2}$ & $1312.84$ &&\\
&&&&&&& $1448.40$ & $3812.06$\\
		\end{tabular}
		\end{ruledtabular}
		\end{table*}
		
\begin{figure*}[tb]  
\centering
\begin{subfigure}{0.45\textwidth}
\includegraphics[height=6cm,width=8.5cm]{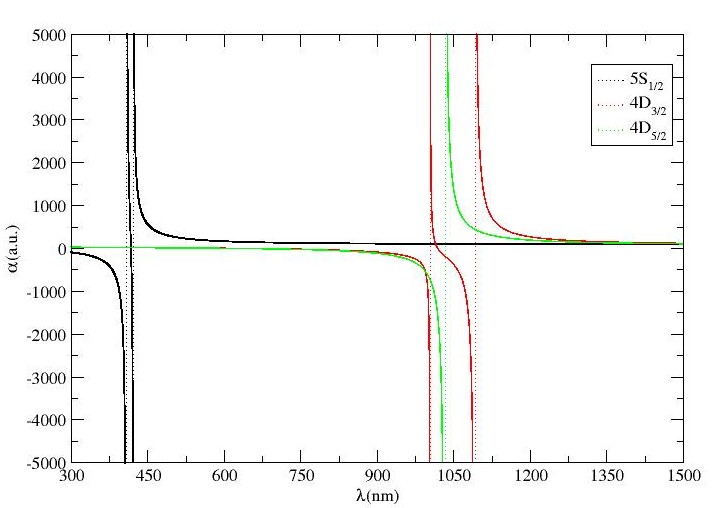}
\caption{$\alpha$ v/s $\lambda$ plot for $5S_{1/2}$ and $4D_{3/2,5/2}$ states of Sr$^{+}$ ion.}
\label{Sr54}
\end{subfigure}
\begin{subfigure}{0.45\textwidth}
\includegraphics[height=6cm,width=8.5cm]{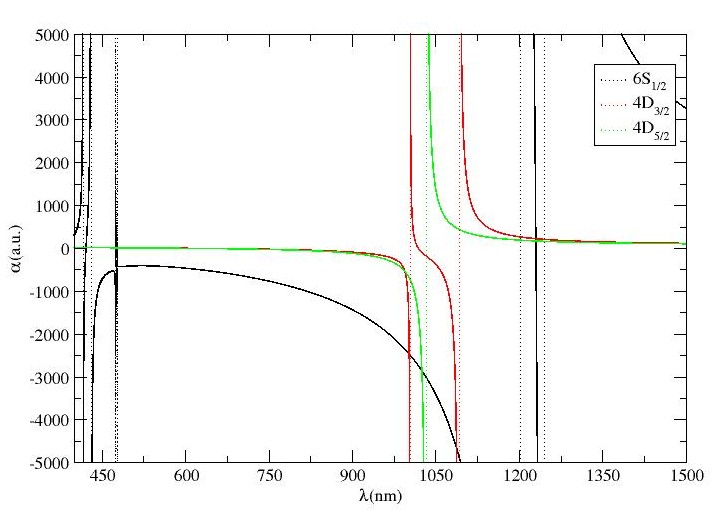}
\caption{$\alpha$ v/s $\lambda$ plot for $6S_{1/2}$--$4D_{3/2,5/2}$ transition in Sr$^{+}$ ion.}
\label{Sr64}
\end{subfigure}
\begin{subfigure}{0.45\textwidth}
\includegraphics[height=6cm,width=8.5cm]{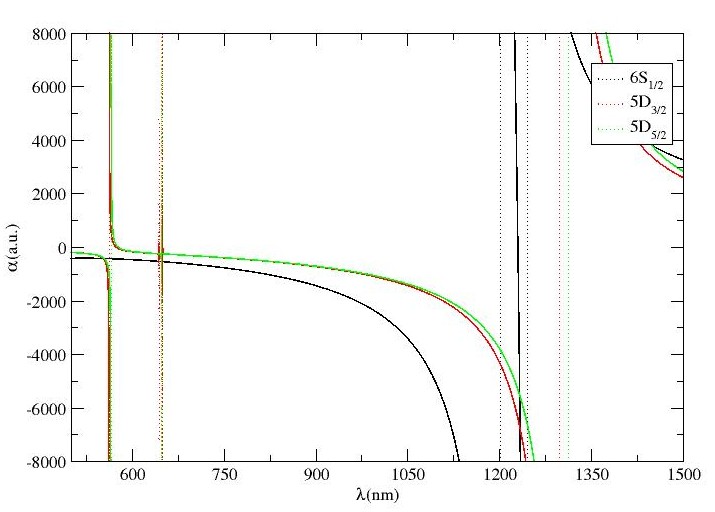}
\caption{$\alpha$ v/s $\lambda$ plot for $6S_{1/2}$--$5D_{3/2,5/2}$ transition in Mg$^{+}$ ion.}
\label{Sr65}
\end{subfigure}
\begin{subfigure}{0.45\textwidth}
\includegraphics[height=6cm,width=8.5cm]{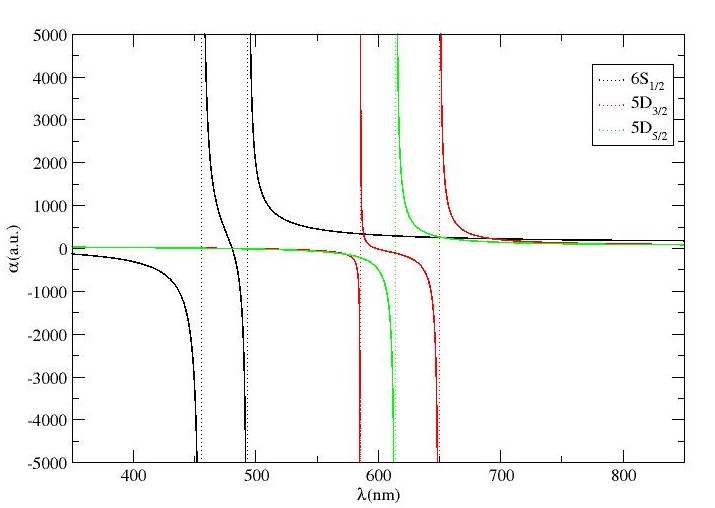}
\caption{$\alpha$ v/s $\lambda$ plot for $6S_{1/2}$--$5D_{3/2,5/2}$ transition in Ba$^{+}$ ion.}
\label{Ba65}
\end{subfigure}
\begin{subfigure}{0.45\textwidth}
\includegraphics[height=6cm,width=8.5cm]{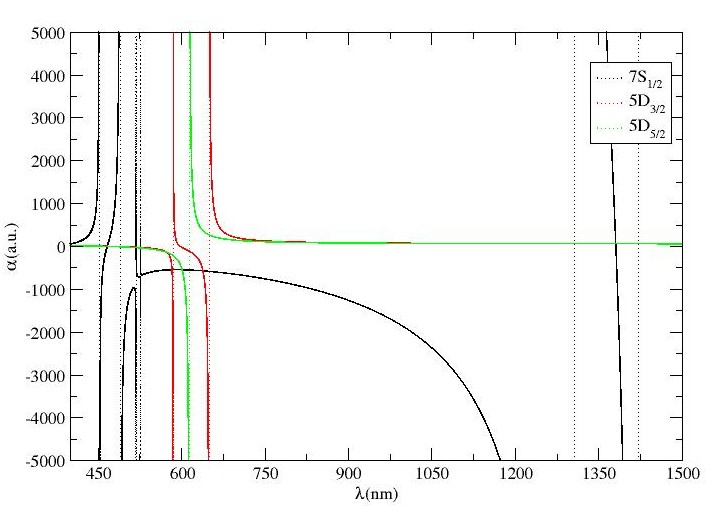}
\caption{$\alpha$ v/s $\lambda$ plot for $7S_{1/2}$--$5D_{3/2,5/2}$ transition in Ba$^{+}$ ion.}
\label{Ba75}
\end{subfigure}
\begin{subfigure}{0.45\textwidth}
\includegraphics[height=6cm,width=8.5cm]{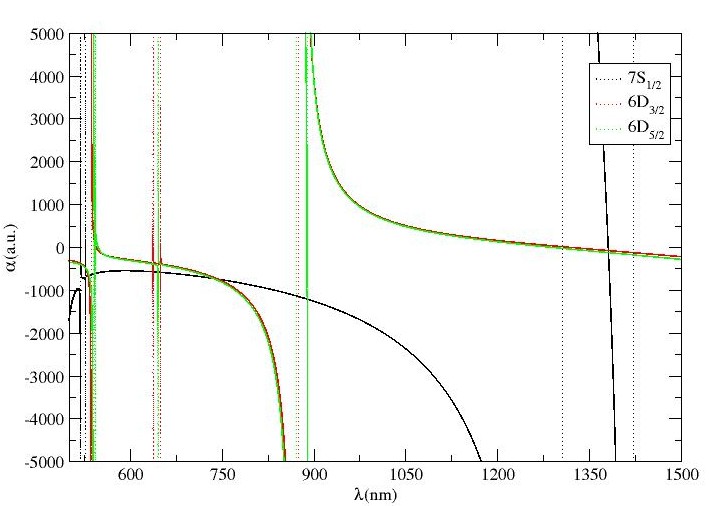}
\caption{$\alpha$ v/s $\lambda$ plot for $7S_{1/2}$--$6D_{3/2,5/2}$ transition in Ba$^{+}$ ion.}
\label{Ba76}
\end{subfigure}
\end{figure*}

Fig.s \ref{Sr54}, \ref{Sr64} and \ref{Sr65} demonstrate the M$_{J}$-independent dynamic dipole polarizability versus wavelength plots for $(6,5)S_{1/2}$--$4D_{3/2,5/2}$ and $6S_{1/2}$--$5D_{3/2,5/2}$ transitions for Sr$^+$ ion. The results corresponding to these figures have been enlisted in Table \ref{table7}. Only two magic wavelengths have been traced for $5S$--$4D_{3/2}$ transition, whereas only one magic wavelength exists for $5S$--$4D_{5/2}$ transition. According to Table \ref{table7}, for $6S$--$4D_{3/2}$ transition, three magic wavelengths exist below $480$ nm, with a dynamic polarizability of value less than $15$ a.u., however, other three $\lambda_{magic}$s, lie between $1000$ nm and $1231$ nm. The $\lambda_{magic}$s at $1002.401$ nm and $1087.35$ nm support blue-detuned traps with sufficiently high polarizabilities for experimental trapping of Sr$^+$ ion. For $6S$--$4D_{5/2}$ transition, five magic wavelengths have been located between $420$ nm and $1250$ nm, out of which, the magic wavelengths at $421.47$ nm, $474.61$ nm, $477.56$ nm and $1239.05$ nm follow red-detuned traps whereas the only magic wavelength at $1025.19$ nm with corresponding $\alpha=-2857.98$ a.u., supports blue-detuned trap which can be useful for experimental purposes. We recommend this magic wavelength of Sr$^+$ ion for $6S$-$4D_{5/2}$ transition. Moreover, it is also observed that all the magic wavelengths for these two transitions lie between same resonance transitions and are closer to each other. So, it is probable to trap Sr$^+$ ion for both of these transitions with same magic wavelength. Table \ref{table7} also shows that there are four magic wavelengths which lie within the wavelength range of $640$ nm to $1450$ nm for $6S$--$5D_{3/2}$ transition. It is also observed that three out of four magic wavelengths for $6S$--$5D_{3/2}$ transition support blue-detuned traps, however the $\lambda_{magic}=1233.61$ nm at $\alpha_{magic}=-6755.64$ a.u. is recommended for experimental purposes as it is far-detuned and a high value of dipole polarizability indicates deep trapping potential. On the other hand, only three magic wavelengths have been identified for $6S$--$5D_{5/2}$ transition in Sr$^+$ ion with two supporting blue-detuned traps. Two out of these $\lambda_{magic}$s, i.e., $1233.06$  nm and $1448.40$ nm are located at higher wavelength range, with deep potentials for their respective favourable blue- and red-detuned traps. Therefore, both of these values are recommended for further experimental studies. Moreover, we have compared our magic wavelengths for $5S$--$4D_{3/2,5/2}$ transitions with respect to available literature in the same table. It is seen that our reported values are in excellent approximation with the results obtained by Kaur et~al.~\cite{kaur2017annexing} with a variation less than $0.05\%$. Unfortunately, we couldn't find any data related to other transitions to carry out the comparison with. Hence, it can be concluded from the comparison of available data that our results are promising and can be used for further prospective calculations of atomic structures and atomic properties of this ion.
	
\subsubsection{Ba$^+$}
\begin{table*}[ht!]
\caption{\label{table10}Magic wavelengths $\lambda_{magic}$ (in nm) with the corresponding polarizability $\alpha_{n}(\omega)$ (in a.u.) along with their comparison with available literature for $6S_{1/2}$--$5D_{3/2,5/2}$ transitions in Ba$^{+}$ ion.
	}
	\begin{ruledtabular}
\begin{tabular}{cccccccccc}
\multicolumn{4}{c}{$6S_{1/2}-5D_{3/2}$}  & & \multicolumn{4}{c}{$6S_{1/2}-5D_{5/2}$}\\
Resonance & $\lambda_{res}$ & $\lambda_{magic}$ & $\alpha_{magic}$ & & Resonance & $\lambda_{res}$ & $\lambda_{magic}$ & $\alpha_{magic}$\\
\hline
$6S_{1/2}\rightarrow 6P_{3/2}$ & $455.53$ & & & & $6S_{1/2}\rightarrow 6P_{3/2}$ & $455.53$ & & &\\
&& $480.710$ & $-4.10$ & & & & $480.76$ & $-8.32$\\
&& $480.6(5)$~\cite{kaur2017annexing} & $-2.89~$\cite{kaur2017annexing} &&&&&\\
$6S_{1/2}\rightarrow 6P_{1/2}$ & $493.55$ & & & & $6S_{1/2}\rightarrow 6P_{1/2}$ & $493.55$
& & \\
$5D_{3/2}\rightarrow 6P_{3/2}$ & $585.53$ & & && $5D_{3/2}\rightarrow 6P_{3/2}$ & $614.34$ &&\\
&& $588.32$ & $330.15$ & & & & $653.17$ & $247.90$\\
&& $588.4(3)$~\cite{kaur2017annexing} & $329.33$~\cite{kaur2017annexing} &&&& $695.7(3)$~\cite{kaur2017annexing} & $219.4$~\cite{kaur2017annexing}\\
$5D_{3/2}\rightarrow 6P_{1/2}$ & $649.87$ & & && & & &&\\
&& $693.46$ & $221.91$ & & & & &\\
&& $655.50(3)$~\cite{kaur2017annexing} & $244.89$~\cite{kaur2017annexing} &&&&\\
&&&&&&&&\\
\hline
\multicolumn{4}{c}{$7S_{1/2}-5D_{3/2}$}  & & \multicolumn{4}{c}{$7S_{1/2}-5D_{5/2}$}\\
\hline
&&&&&&&&\\
$6P_{1/2}\rightarrow 7S_{1/2}$ & $452.62$ & & & & $6P_{1/2}\rightarrow 7S_{3/2}$ & $452.62$ & & &\\
&& $466.952$ & $0.52$ & & & & $466.883$ & $-2.63$\\
$6P_{3/2}\rightarrow 7S_{1/2}$ & $490.13$ & & & & $6P_{3/2}\rightarrow 7S_{1/2}$ & $490.13$
& & \\
$7S_{1/2}\rightarrow 8P_{3/2}$ & $518.49$ & & && $7S_{1/2}\rightarrow 8P_{3/2}$ & $518.49$ &&\\
&& $518.79$ & $-22.85$ & & & & $518.79$ & $-31.91$\\
$7S_{1/2}\rightarrow 8P_{1/2}$ & $526.75$ & & && $7S_{1/2}\rightarrow 8P_{1/2}$ & $526.75$ &&\\
&& $526.78$ & $-28.72$ & & & & $601.37$ & $-548.78$\\
&& $583.76$ & $-548.592$ & & & & &\\
$5D_{3/2}\rightarrow 6P_{3/2}$ & $585.53$ & & && $5D_{5/2}\rightarrow 6P_{3/2}$ & $614.34$ &&\\
&& $638.75$ & $-573.91$ & & & & &\\
$5D_{3/2}\rightarrow 6P_{1/2}$ & $649.87$ & & && & & &&\\
$7S_{1/2}\rightarrow 7P_{3/2}$ & $1306.14$ & & & & $7S_{1/2}\rightarrow 7P_{3/2}$ & $1306.14$ &&\\
&& $1380.83$ & $59.75$ & & & & $1380.83$ & $59.15$\\
$7S_{1/2}\rightarrow 7P_{1/2}$ & $1421.54$ & & && $7S_{1/2}\rightarrow 7P_{1/2}$ & $1421.54$& &\\
&&&&&&&&\\
\hline
\multicolumn{4}{c}{$7S_{1/2}-6D_{3/2}$}  & & \multicolumn{4}{c}{$7S_{1/2}-6D_{5/2}$}\\
\hline
&&&&&&&&\\
$7S_{1/2}\rightarrow 8P_{3/2}$ & $518.49$ & & & & $7S_{1/2}\rightarrow 8P_{3/2}$ & $518.49$ & & &\\
&& $519.03$ & $-375.84$ & & & & $519.07$ & $-405.03$\\
$7S_{1/2}\rightarrow 8P_{1/2}$ & $526.750$ & & & & $7S_{1/2}\rightarrow 8P_{1/2}$ & $526.75$ & & \\
&& $526.84$ & $-484.17$ & & & & $526.839$ & $-479.94$\\
&& $530.98$ & $-664.18$ & & & & $532.797$ & $-655.19$\\
$6D_{3/2}\rightarrow 6F_{5/2}$ & $536.28$ & & && $6D_{5/2}\rightarrow 6F_{7/2}$ & $539.31$ &&\\
&&&&&&& $542.17$ & $-613.07$\\ 
&&&&& $6D_{5/2}\rightarrow 6F_{5/2}$ & $542.26$ &&\\ 
&&&&&&& $645.33$ & $-580.80$\\ 
$6D_{3/2}\rightarrow 8P_{3/2}$ & $637.25$ & & && $6D_{5/2}\rightarrow 8P_{3/2}$ & $645.70$ &&\\
&&&&&&& $735.65$ & $-728.27$\\ 
$6D_{3/2}\rightarrow 8P_{1/2}$ & $649.77$ & & && $6D_{5/2}\rightarrow 5F_{7/2}$ & $871.32$ &&\\
&& $743.97$ & $-746.39$ & & & & $889.20$ & $-1211.62$\\
$6D_{3/2}\rightarrow 5F_{5/2}$ & $874.02$ & & & & $6D_{5/2}\rightarrow 5F_{5/2}$ & $889.99$ &&\\
$7S_{1/2}\rightarrow 7P_{3/2}$ & $1306.14$ & & && $7S_{1/2}\rightarrow 7P_{3/2}$ & $1306.14$& &\\
&& $1381.25$ & $-77.60$ & & & & $1381.39$ & $-125.03$\\
$7S_{1/2}\rightarrow 7P_{1/2}$ & $1421.54$ & & & & $7S_{1/2}\rightarrow 7P_{1/2}$ & $1421.54$ &&\\
		\end{tabular}
		\end{ruledtabular}
		\end{table*}
		
The results for magic wavelengths for $6S$--$5D_{3/2,5/2}$, $7S$--$5D_{3/2,5/2}$ and $7S$--$6D_{3/2,5/2}$ transitions in Ba$^+$ ion are tabulated in tables \ref{table10}. As per Fig. \ref{Ba65} and Table\ref{table10}, A maximum of magic wavelengths have been located between $480$ and $700$ nm. It is also observed that the magic wavelengths that lie between $6S$--$6P_{1/2}$ and $6S$--$6P_{3/2}$ resonant transitions support blue-detuned trap, however, the dynamic dipole polarizability corresponding to these magic wavelengths are too small to trap Ba$^+$ ion at these wavelengths. A total of six magic wavelengths are found for $7S$--$5D_{3/2}$ transition out of which two lie in the vicinity of $526$ nm. The sharp intersection of polarizability curves of the involved states of transition lie at $583.76$ nm, $638.75$ nm and $1380.83$ nm. Similarly, four magic wavelengths have been identified for $7S$--$5D_{5/2}$ transition, however, unlike $7S$-$5D_{3/2}$ transition, no magic wavelength has been identified in the vicinity of $600$ to $1300$ nm. It is also analyzed that three out of these four $\lambda_{magic}$s, support blue-detuned trap, although the trapping potentials for these traps are not deep enough for further consideration to experimentations.
		
Table\ref{table10} also compiles the magic wavelengths for $7S$--$5D_{3/2,5/2}$ transitions and shows that there exists six and four magic wavelengths for $7S$--$5D_{3/2}$ and $7S$--$5D_{5/2}$ transition, respectively. It is also seen that the magic wavelengths between $6P_{3/2}$--$7S$ and $7S$--$8P_{3/2}$ as well as $5D_{3/2}$--$6P_{1/2}$ and $7S$--$7P_{3/2}$ transitions seem to be missing as shown in Fig. \ref{Ba75}. It is also observed that the magic wavelength at $466.95$ nm and $1380.83$ nm are slightly red-shifted, nevertheless, the $\lambda_{magic}$ at $638.75$ nm lies in visible region supports blue-detuned trap, can have sufficient trap depth at reasonable laser power.

Similarly, the magic wavelengths and their corresponding dynamic dipole polarizability along with their comparison with available literature is also provided in the same table for $7S$--$6D_{3/2,5/2}$ transitions. The same have been demonstrated graphically in the Fig. \ref{Ba76} which includes a total of thirteen magic wavelengths in all for the considered transitions. It is also examined that no magic wavelength exists between $6D_{3/2}$--$6F_{5/2}$ and $6D_{3/2}$--$8P_{1/2}$ resonances. Unlike $7S$--$6D_{3/2}$ transition, around eight magic wavelengths have been located between $7S$--$8P_{3/2}$ and $7S$--$7P_{1/2}$ resonances, and all of them support blue-detuned traps. Moreover, magic wavelengths at $532.80$ nm, $735.65$ nm and $1381.39$ nm are expected to be more promising for experiments due to sufficient trap depths for the reasonable power lasers. However, on the comparison of our results for $6S$--$5D_{3/2,5/2}$ transitions for Ba$^+$ ion, we have observed that all the magic wavelengths agree well with the results obtained by Kaur et~al. in Ref.~\cite{kaur2017annexing}, except the last magic wavelengths that are identified at $693$ nm and $653$ nm for $6S$--$5D_{3/2}$ and $6S$--$5D_{5/2}$ transitions. 

\subsection{\label{tune}Tune-out Wavelengths}
\begin{table*}[ht!]
	\caption{\label{tabletune}%
		Tune-out wavelengths $\lambda_{T}$ (in nm) various states of Mg$^+$, Ca$^+$, Sr$^+$ and Ba$^+$ ions and their comparison with available literature.
	}
	\begin{ruledtabular}
\begin{tabular}{ccccccccccccc}
\multicolumn{3}{c}{Mg$^+$} & \multicolumn{3}{c}{Ca$^+$} &  \multicolumn{3}{c}{Sr$^+$} & \multicolumn{3}{c}{Ba$^+$}\\ 
State & $\lambda_{T}$ & Others & State & $\lambda_{T}$ & Others & State & $\lambda_{T}$ & Others & State & $\lambda_{T}$ & Others \\
\hline
$3S_{1/2}$ & $102.61$ & & $4S_{1/2}$ & $165.04$ & & $5S_{1/2}$ & $417.04$ & $417.04(6)$~\cite{kaur2021tune} & $6S_{1/2}$ & $200.04$ & \\
 & & & & & & & & $417.025$~\cite{kaur2017annexing} & & &\\
 & $102.70$ & & & $165.26$ & & & & & & $202.47$ & \\
 & $124.02$ & & & $395.80$ & $395.80(2)$~\cite{kaur2021tune} & & & & & $202.51$ &\\
 & & & & & $395.796$~\cite{kaur2017annexing} & & & & & &\\
 & $280.11$ & $280.110(9)$~\cite{kaur2021tune} & & & & & & & & $480.663$ & $480.63(24)$~\cite{kaur2021tune}\\
& & & & & & & & & & & $480.66(18)$~\cite{jiang2021tune}\\ 
& & & & & & & & & & & $480.596$~\cite{kaur2017annexing}\\
$3D_{3/2}$ & $317.08$ & & $3D_{3/2}$ & $212.93$ & & $4D_{3/2}$ & $185.50$ & & $5D_{3/2}$ & $224.68$ &\\
 & $384.96$ & & & $213.25$ & & & $192.95$ & & & $468.61$ & $472.461$~\cite{kaur2017annexing}\\
 & $385.34$ & & & $494.37$ & $492.752$~\cite{kaur2017annexing} & & $242.62$ & & & $597.93$ & $597.983$~\cite{kaur2017annexing}\\
 & $812.03$ & & & $852.75$ & $852.776$~\cite{kaur2017annexing} & & $606.47$ & $598.633$~\cite{kaur2017annexing} & & &\\
 & {$1092.44$} & & & & & & $1018.911$ & $1018.873$~\cite{kaur2017annexing} & & &\\
$3D_{5/2}$ & $317.00$ & & $3D_{5/2}$ & $170.63$ & & $4D_{5/2}$ & $193.78$ & & $5D_{5/2}$ & $193.32$ &\\
 & $385.14$ & & & $213.16$ & & &  $242.56$ & & & $198.43$ &\\
 & $810.70$ & & & $493.13$ & $482.642$~\cite{kaur2017annexing} & & $594.03$ & $585.677$~\cite{kaur2017annexing}& & $225.63$ &\\
 & & & & & & & & & & $234.77$ &\\
 & & & & & & & & & & $459.57$ & $509.687$~\cite{kaur2017annexing}\\
$4S_{1/2}$ & $279.13$ & & $5S_{1/2}$ & $287.18$ & & $6S_{1/2}$ & $334.78$ & & $7S_{1/2}$ & $347.629$ &\\
 & $293.21$ & & & $299.64$ & & & $362.58$ & & & $363.36$ &\\
 & $361.52$ & & & $309.13$ & & & $378.37$ & & & $394.63$ &\\
 & $361.69$ & & & $309.30$ & & & $421.40$ & & & $466.94$ &\\
 & & & & $371.75$ & & & $474.62$ & & & $518.78$ &\\
 & & & & $447.39$ & & & $477.56$ & & & $526.78$ &\\
 & & & & $448.09$ & & & $1230.15$ & & & $1381.01$ &\\
 & & & & $1191.59$ & & & & & & &\\
$4D_{3/2}$ & $554.11$ & & $4D_{3/2}$ & $317.85$ & & $5D_{3/2}$ & $407.44$ & & $6D_{3/2}$ & $544.8$ &\\
 & $592.70$ & & & $334.65$ & & & $434.69$ & & & $1316.93$ &\\
 & $1331.53$ & & & $371.75$ & & & $481.52$ & & & &\\
 & & & & $375.78$ & & & $573.81$ & & & &\\
 & & & & $471.96$ & & & $649.82$ & & & &\\
$4D_{5/2}$ & $507.25$ & & $4D_{5/2}$ & $472.38$ & & $5D_{5/2}$ & $436.25$ & & $6D_{5/2}$ & $547.53$ &\\
 & $553.95$ & & & $376.06$ & & & $455.83$ & & & $889.04$ &\\ 
 & $592.55$ & & & & & & $577.18$ & & & $1287.17$ &\\
 & $811.97$ & & & & & & $647.60$ & & & &\\
 & $1329.45$ & & & & & & {$1312.630$} & & & &\\
    
		\end{tabular}
		\end{ruledtabular}
		\end{table*}		
We have illustrated tune-out wavelengths for different states of the considered transitions in the alkaline-earth ions along with their comparison with already available literature in Table \ref{tabletune}. To locate these M$_J$-independent tune-out wavelengths, we have evaluated scalar dipole dynamic polarizabilities of these states for considered alkaline-earth ions and identified those values of $\lambda$ for which polarizability vanished. It is also accentuated that in Mg$^+$ ion, all the tune-out wavelengths identified for $3S_{1/2}$ and $4S_{1/2}$ states lie in UV region, whereas for $(3,4)D_{3/2,5/2}$ states, a few tune-out wavelengths are located in visible range. Moreover, the largest $\lambda_{T}$ is identified for $4D_{3/2}$ state at $1331.527$ nm. Furthermore, only one tune-out wavelength,i.e., $\lambda_T=280.11$ nm for $3S_{1/2}$ could be compared with the result presented by Kaur et~al. in Ref.~\cite{kaur2021tune} and it is seen that our result is in good accord with this value. 
Similarly, we have pointed out tune-out wavelengths for $nS_{1/2}$ and $(n-1)D_{3/2}$, $n=(4,5),(5,6)$ and $(6,7)$ states for Ca$^+$, Sr$^+$ and Ba$^+$ ions, by identifying $\lambda$s at which their corresponding $\alpha$s tend to zero. Hence, it has been perceived that out of $25$ tune-out wavelengths for all states of Ca$^+$ ion, only seven of them lie within visible spectrum and on comparison of different tune-out wavelengths for $4S_{1/2}$ and $3D_{3/2}$ states of Ca$^+$ ion, it has been analyzed that all of these results are advocated by the results obtained in Refs.~\cite{kaur2017annexing, kaur2021tune}. {However, one of the tune-out wavelength has been located at $493.13$ nm for $3D_{5/2}$ state of Ca$^+$ ion seems have $2\%$ variation from the wavelength obtained by Kaur et~al. in Ref.~\cite{kaur2017annexing}. This may be due to the fact that our study incorporates all the highly precise E1 matrix elements as well energies of the states available at Portal for High-Precision Atomic Data and Computation~\cite{UDportal}, which appears to be missing in previous studies.}
For Sr$^+$ ion, maximum number of tune-out wavelengths have been identified out of all the considered alkaline-earth ions. It is also realized that most of these $\lambda$s lie within visible spectrum of electromagnetic radiation, are mostly comprise of all the $\lambda_T$ values corresponding to $5S_{1/2}$, $5D_{3/2}$ and $5D_{5/2}$ states. Additionally, during the comparison of these values with the results published in Refs.~\cite{kaur2017annexing,kaur2021tune}, it is examined that tune-out wavelength at $417.04$ nm for $5S_{1/2}$ as well as $\lambda_T=1018.91$ nm for $4D_{3/2}$ state agree well with the available results, howsoever, the tune-out wavelengths at $606.50$ nm and $594.03$ nm for $4D_{3/2}$ and $4D_{5/2}$ states, respectively show a discrepancy of less than $2\%$ which lies within quoted error limit. 
In case of Ba$^+$ ion, we have located $24$ tune-out wavelengths, which in all comprise of $10$, $10$ and $4$ wavelengths in visible, UV and infrared regions, respectively. It is also accentuated that all the tune-out wavelengths that exist in visible region lie within the range $480$ nm to $550$ nm. We have also compared our tune-out wavelengths for $6S_{1/2}$ and $5D_{3/2,5/2}$ states against available theoretical data in Refs.~\cite{kaur2017annexing,kaur2021tune,jiang2021tune} and it is found that all the $\lambda_T$s except $468.61$ nm and $459.570$ nm, respectively for $5D_{3/2}$ and $5D_{5/2}$ states show disparity less than $1\%$ which lies within the considerable error limit.

\section{\label{4} Conclusion}

We have identified a number of reliable magnetic-sublevel independent tune-out wavelengths of many $S_{1/2}$ and $D_{3/2,5/2}$ states, and magic 
wavelengths of different combinations of $S_{1/2}$--$D_{3/2,5/2}$ transitions in the alkaline-earth ions from Mg$^+$ through Ba$^+$. If they 
can be measured precisely, accurate values of many electric dipole matrix elements can be inferred by combining the experimental values of these 
quantities with our theoretical results. Most of the magic wavelengths found from this study show that they can be detected using the red and 
blue-detuned traps. In fact, it is possible to perform many high-precision measurements by trapping the atoms at the reported tune-out and magic
wavelengths of the considered transitions in the future that can be applied to different metrological studies.

\bibliographystyle{unsrt}

\bibliography{Magsub.bib}

\end{document}